\documentclass[useAMS,usegraphicx,usenatbib]{mn2e}

\usepackage{graphicx}
\usepackage{subfigure}
\usepackage{color}
\usepackage{amsmath}
\usepackage{amssymb}
\usepackage{mathptmx}
\usepackage{times}
\usepackage{epsfig}
\usepackage{mathrsfs}
\usepackage{epstopdf}

\usepackage{color}

\usepackage[colorlinks]{hyperref}
\hypersetup{ colorlinks, linkcolor=blue, citecolor=blue }

\usepackage{natbib}
\bibpunct{(}{)}{;}{a}{}{,}

\DeclareMathAlphabet{\mathsc}{OT1}{cmr}{m}{sc}
\def\testbx{bx}%
\DeclareRobustCommand{\ion}[2]{%
\relax\ifmmode
\ifx\testbx\f@series
{\mathbf{#1\,\mathsc{#2}}}\else
{\mathrm{#1\,\mathsc{#2}}}\fi
\else\textup{#1\,{\mdseries\textsc{#2}}}%
\fi}

\newcommand{\Nai}{\ion{Na}{i}}
\newcommand{\Feii} {\ion{Fe}{ii}}
\newcommand{\Caii} {\ion{Ca}{ii}}
\newcommand{\Crii} {\ion{Cr}{ii}}

\newcommand{\Feiii} {\ion{Fe}{iii}}

\newcommand{\SiII} {\ion{Si}{ii}}

\newcommand{\Coii}{\ion{Co}{ii}}

\title[A 2002cx-like SN Iax SN 2013en]{Optical observations of a SN~2002cx-like peculiar supernova SN~2013en in UGC~11369}
\author[Z. W. Liu]{Zheng-Wei Liu$^{1}$\thanks{E-mail:zwliu@ynao.ac.cn},
                    J.-J. Zhang$^{2,3}$, F. Ciabattari$^{4}$, L. Tomasella$^{5}$, X.-F. Wang$^{6}$,\newauthor X.-L. Zhao$^{6}$, T.-M. Zhang$^{7,8}$, Y.-X. Xin$^{2,3}$, C.-J. Wang$^{2,3}$ and L. Chang$^{2,3}$\\
$^{1}$Argelander-Institut f\"ur Astronomie, Auf dem H\"ugel 71, D-53121, Bonn, Germany\\
$^{2}$Yunnan Observatories, Chinese Academy of Sciences (CAS), Kunming 650011, P.R. China\\
$^{3}$Key Laboratory for the Structure and Evolution of Celestial Object, CAS, Kunming 650011, P.R. China \\
$^{4}$Monte Agliale Observatory, Borgo a Mozzano, Lucca, 55023, Italy\\
$^{5}$INAF, Osservatorio Astronomico di Padova, 35122 Padova, Italy\\
$^{6}$Physics Department and Tsinghua Center for Astrophysics (THCA), Tsinghua University, Beijing 100084, China\\
$^{7}$National Astronomical Observatories of China (NAOC), Chinese Academy of Sciences, Beijing 100012, China\\
$^{8}$Key Laboratory of Optical Astronomy, National Astronomical Observatories, Chinese Academy of Sciences, Beijing 100012, China\\
}

\begin{document}



\maketitle

\label{firstpage}

\begin{abstract}

We present optical observations of a SN~2002cx-like supernova SN~2013en in 
UGC~11369, spanning from a phase near maximum light ($t=\rm{+1\,d}$) to $t=\rm{+60\,d}$ 
with respect to the $R$-band maximum.  Adopting a distance modulus of $\mu=34.11\pm 0.15\,\rm{mag}$ 
and a total extinction (host galaxy+Milky Way) of $A_{\rm{V}}\approx 1.5\,\rm{mag}$,
we found that SN 2013en peaked at $M_{R}\approx-18.6\,\rm{mag}$, which is underluminous compared to  the normal SNe Ia.   
The  near  maximum spectra  show  lines  of  \SiII, \Feii, \Feiii, \Crii, \Caii\ and other 
intermediate-mass and iron group elements which all have lower expansion velocities (i.e., $\sim\,6000\,\rm{km\,s^{-1}}$).   
The photometric and spectroscopic evolution of SN~2013en is remarkably similar to those of SN~2002cx and SN~2005hk, suggesting that 
they are likely to be generated from a similar progenitor scenario or explosion mechanism. 
 
\end{abstract}

\begin{keywords}
          galaxies:individual:(UGC\,11369) -- supernovae:general -- supernovae:ind-\\ividual:(SN\,2013en)
\end{keywords}

\section{Introduction}

Type Ia supernovae (SNe Ia) have been successfully used as standardizable candles to measure expansion rate of 
the Universe \citep{Ries98, Perl99}. Most SNe Ia (about $70\%$) belong to ``normal'' 
type \citep{Bran93, Li03}, 
they can be calibrated by an empirical relation between their peak luminosities and 
light-curve shapes \citep{Phil93}. However, a new peculiar sub-class of SNe Ia, named type Iax 
supernovae (SNe Iax, see \citealt{Fole13}) after its prototypical member SN~2002cx \citep{Li03}, 
has been discovered to deviate significantly from this relation. SNe Iax are suggested to arise from 
thermonuclear explosions of carbon-oxygen white dwarfs (CO~WDs) because their maximum-light 
spectra seem to show signs of CO burning as in normal SNe Ia \citep{Fole13}. Nevertheless,  
a core-collapse scenario has been suggested for at least one such type of explosion, i.e., SN 2008ha \citep{Vale08, Fole10, Mori10}.

SNe Iax have several observational similarities to normal SNe Ia, but 
also present sufficiently distinct observational properties:
(i)\,their ejecta are dominated by intermediate-mass elements (IMEs) 
and iron-group elements \citep{Fole13}. Also, strong mixing with both IMEs and iron-group elements 
is seen in all layers of the ejecta \citep{Jha06, Phil07, Fole13}. These features are in clear contrast to normal
SNe Ia, which are characterized by strongly layered ejecta \citep{Mazz07}.
 (ii)\,Comparing to normal SNe Ia, SNe Iax are significantly 
fainter (\citealt{Fole13, Fole14}). SNe Iax have a wide range in explosion energies ($10^{49}$--$10^{51}\,\rm{erg}$), 
ejecta masses ($0.15$--$0.5\,\rm{M_{\odot}}$), and $^{56}\rm{Ni}$ masses ($0.003$--$0.3\,\rm{M_{\odot}}$).
(iii)\,The spectra of SNe Iax are characterized by lower expansion velocities ($2000$--$8000\,\rm{km\,s^{-1}}$) 
than those of normal SNe Ia ($\approx$\,$1$--$2\times10^{4}\,\rm{km\,s^{-1}}$) at similar epochs. (iv)\,Instead of entering 
a nebular phase dominated by broad forbidden lines of iron-peak elements in normal SNe Ia, the late-time spectra of 
SNe Iax are dominated by narrow permitted Fe\,II \citep{Jha06}. (v)\,Two SNe Iax (SN 2004cs and SN 2007J) were 
identified  with strong He lines in their spectra \citep{Fole09, Fole13}. However, no He lines have yet 
been detected in spectra of normal SNe Ia.

\citet{Li11} suggested SN~2002cx (or SNe Iax) are perhaps the most common type of peculiar SNe Ia (see also \citealt{Fole13}).
It has been estimated that SNe Iax can contribute about $5$--$30\%$ of the total SN 
Ia rate \citep{Li11, Fole13, Whit14}. By studying the properties of the host galaxies of SNe Iax, it is found that the host-galaxy 
morphology distribution of 02cx-like SNe Ia is highly skewed to late-type galaxies, with none of the present sample 
of 02cx-like SNe Ia occurring in an elliptical 
galaxy \citep{Fole13, Lyma13, Whit14}. However, at least one SN Iax event (SN~2008ge) is hosted by a S0 
galaxy with no signs of star formation. Recently, by analyzing pre- and post-explosion $\textit{HST}$ images of some
SNe Iax \citep{McCu14, Fole14, Fole15} and considering the He lines seen in two SN Iax spectra, a 
relatively young age was suggested  ($<$\,$500\,\rm{Myr}$, see \citealt{Fole14}) for the progenitor systems of SNe Iax.

\begin{figure}
\centering
\includegraphics[width=0.48\textwidth, angle=360]{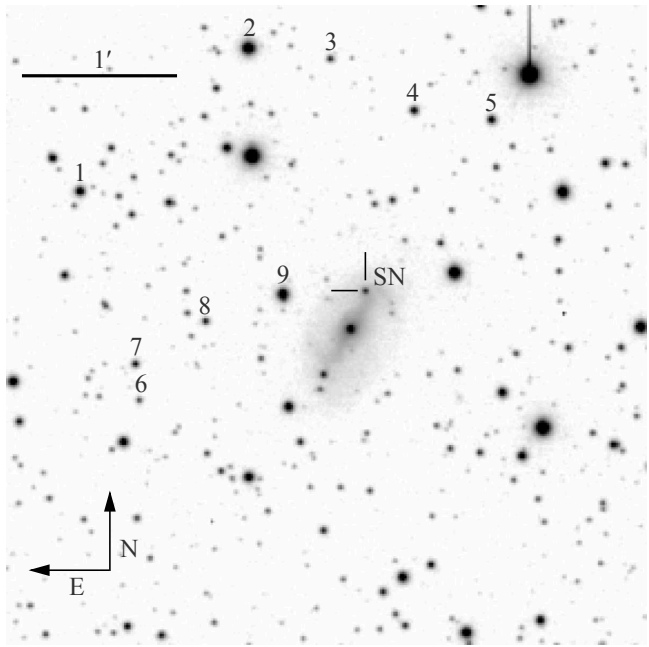}
\caption{The LJT $R$-band image of SN 2013en and its reference
stars, taken on Aug. 3.61, 2013. The image scale is 0$\arcsec$.283/pixel.
Nine local standard stars are labeled with Arabic numerals.
         }
\label{Fig:1}
\end{figure}

Despite recent progress on both, the theoretical and observational side (e.g., see \citealt{Bran04, 
Jha06, Phil07, Fole09, Vale08, Vale09, Mori10, Shen10, Fole12, Fole13, Krom13, Stri14, Liu13, Liu15, Krom15}), the problem 
of progenitors of SNe Iax is still poorly constrained.  Theoretically, the hydrodynamical simulations have shown that weak deflagrations 
of the Chandrasekhar-mass (Ch-mass) CO WDs \citep{Jord12, Krom13, Fink13} and Ch-mass hybrid WDs \citep{Krom15} seem to provide a viable physical scenario for SNe Iax.
Observationally, a possible progenitor system of one SN Iax event, SN 2012Z, was first detected 
by \citet{McCu14}. It is further suggested that the SN 2012Z likely had a progenitor system which contained a non-degenerate He-star companion
\citep{McCu14, Stri14}. 
However, the same analyses for pre-explosion $\textit{HST}$ images of other SNe Iax seem to indicate 
that SNe Iax must have a diverse set of progenitor systems \citep{Fole14, Fole15}. Future observations 
providing a bigger sample of SNe Iax and more detailed modeling for various theoretical progenitor scenarios will 
be very helpful for constraining the nature of SN Iax progenitors. To date, no single published model has been found to be 
able to explain all the observational features and full diversity of SNe Iax, a combination of 
several progenitor scenarios might be an alternative way to explain the properties of the different observed 
events within SNe Iax.

\begin{figure}
\centering
\includegraphics[width=0.48\textwidth, angle=360]{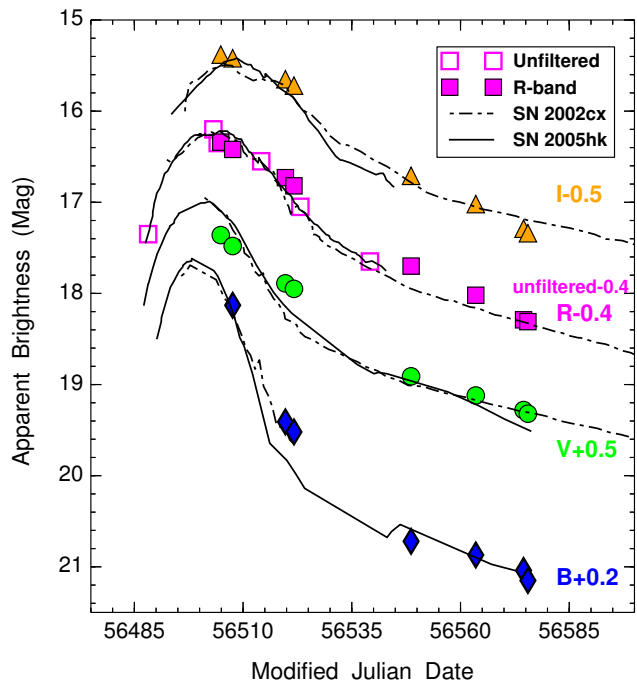}
\caption{$\textit{BVRI}$ (filled markers) and unfiltered (open squares) light curves of SN 20013en. The LCs have been shifted by the amount 
         indicated in the legend. For a comparison, the light curves of SNe 2002cx ($\textit{BVRI}$ band, dash-dotted lines) and 
         2005hk ($\textit{BVRI}$ band, solid lines) are also shown. The light curves of SNe 2002cx and 2005hk have been shifted
         to match the light curve of SN 2013en at peak. Here, the uncertainties for data points are smaller than the plotted symbols.   
         }
\label{Fig:2}
\end{figure}

\begin{table}
\centering
\caption{Magnitudes of the Photometric Standards in the Field of SN 2013en.}
 \begin{tabular}{@{}cccccc@{}}
 \hline\hline
Star & $B$(mag) & $V$(mag) & $R$(mag) & $I$(mag) \\
\hline
1	&	16.150(005)	&	15.417(003)	&	14.966(004)	&	14.646(002)	\\
2	&	15.310(003)	&	13.887(002)	&	13.093(002)	&	12.454(001)	\\
3	&	18.219(025)	&	17.175(012)	&	16.569(008)	&	16.052(007)	\\
4	&	17.577(015)	&	16.329(006)	&	15.591(005)	&	14.989(003)	\\
5	&	16.824(008)	&	16.027(005)	&	15.542(005)	&	15.043(003)	\\
6	&	18.636(030)	&	17.739(015)	&	17.206(010)	&	16.785(008)	\\
7	&	17.672(018)	&	16.781(008)	&	16.274(006)	&	15.838(006)	\\
8	&	18.376(025)	&	17.187(012)	&	16.511(008)	&	15.988(006)	\\
9	&	15.545(004)	&	14.448(002)	&	13.804(002)	&	13.204(002)	\\
\hline
\end{tabular}

\medskip
\flushleft
\textbf{Note.} (1)The finder chart of SN~2013en and nine comparison local stars are shown in Fig.~\ref{Fig:1}.
(2)Uncertainties, in units of 0.01 mag, are 1$\sigma$.
\label{Tab:Ref}
\end{table}

In this work, we present optical observations and analysis of a 2002cx-like SN~2013en. The paper is organized as follows. Observations and
data reductions are described in Section~\ref{sec:data}. In Section~\ref{sec:analysis} we performed the analysis of the 
light curves and spectra of SN~2013en, respectively. A comparison with SN~2002cx, SN~2005hk, and other SNe Ia are also presented 
in this section. A simple discussion about the possible progenitor systems of SNe Iax and a summary 
of the main results of this paper are given in Section~\ref{sec:discussion}.

\begin{figure}
\centering
\includegraphics[width=0.48\textwidth, angle=360]{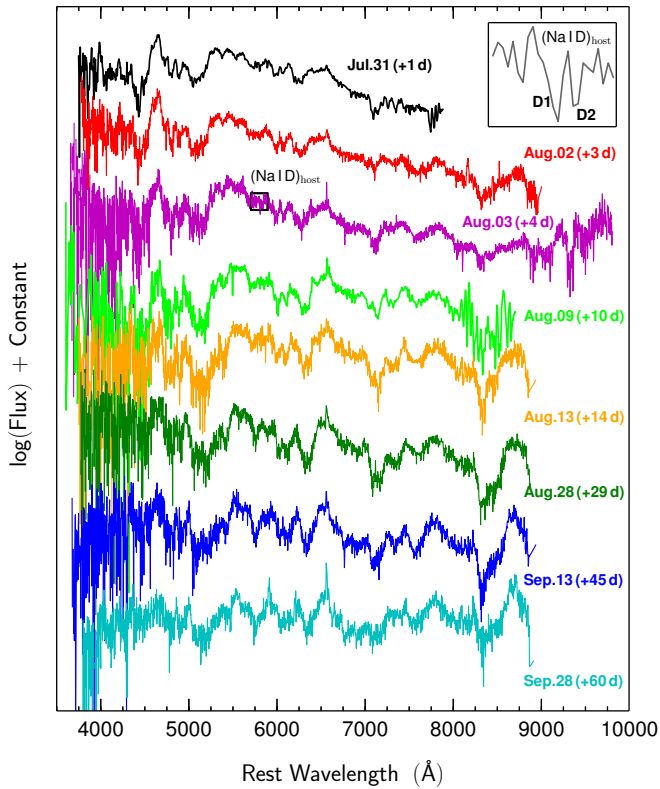}
\caption{Time evolution of the optical spectra of SN 2013en.  The spectra have been 
         corrected for the redshift of the host galaxy. The phases indicated to the right of 
         each spectrum are relative to the $R$-band maximum light. The \Nai\ D absorption 
        of the host galaxy in the spectrum at $t\approx+4$\,days is marked with black 
        box and amplified on the top right.
         }
\label{Fig:3}
\end{figure}

\section{Observations and Data reduction}

\label{sec:data}

SN 2013en (= PSN J18513735+2338206) in UGC 11369  was discovered \citep{Ciab13} 
in an unfiltered CCD image (limiting magnitude of $\rm{19.5\,mag}$) taken by 
a 0.5 meter Newtonian telescope in the course of the Italian Supernovae 
Search Project (ISSP)\footnote{http://italiansupernovae.org} on July 30.98 UT. Its J2000.0 coordinates are $\rm{R.A. = 18^{h}51^{m}37^{s}.35}$, 
$\rm{Decl. = +23^{d}{38}'{20}''.6}$, located at ${6}''$ west and ${17}''$ north 
of the center of the galaxy UGC 11369 (see Figure~\ref{Fig:1}).

\begin{figure}
\centering
\includegraphics[width=0.48\textwidth, angle=360]{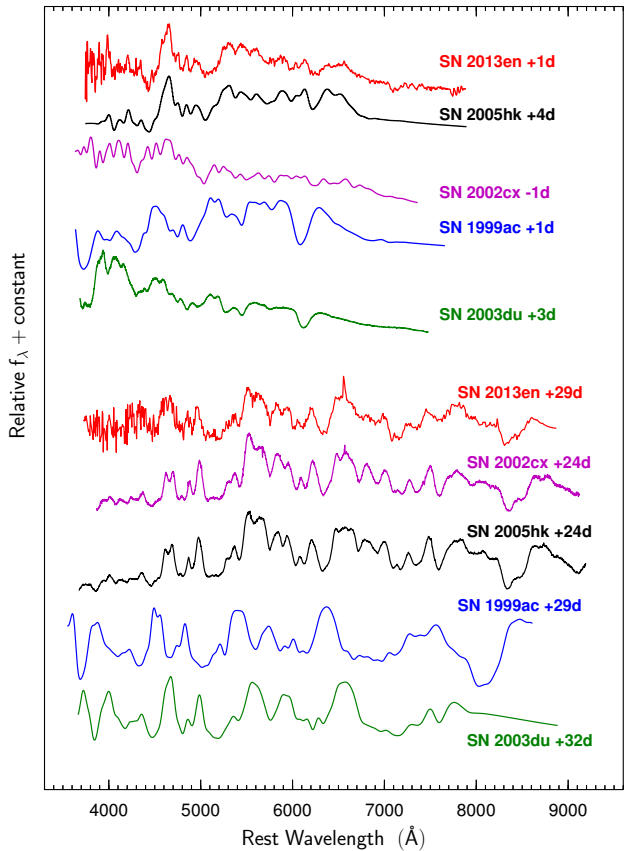}
\caption{Comparison of spectra of SN~2013en at +1, and +29 days with SN~2002cx (magenta curve), SN~2005hk (black curves),
         a  peculiar, 1991T-like SN Ia (SN~1999ac; blue), and normal SNe Ia (SN~2003du;
         green) at similar epoch.
         }
\label{Fig:4}
\end{figure}

Shortly after the discovery, SN~2013en was classified as a peculiar SN Ia by Padova-Asiago Supernova Group \citep{Toma14, Ciab13} from an
optical spectrum (range 340--820\,nm) obtained on July 31.84\,UT with
the Asiago 1.82-m Copernico Telescope (+AFOSC).\footnote{http://sngroup.oapd.inaf.it}
With the supernova identification tool of {\sc GELATO} \citep{Haru08} and {\sc SNID} \citep{Blon07}, a good match between
SN~2013en and a typical SN Iax SN~2005hk is found by adopting the redshift $z=0.015207$ of the 
host galaxy of SN~2013en.

Most optical photometry of SN~2013en was obtained in broad $BVRI$ bands  with the Yunnan Faint Object 
Spectrograph and Camera (YFOSC,  \citealt{Zhang14})  on Li Jiang 2.4-m Telescope (LJT) of Yunnan 
observatories (YNAO), which started on 2013 Aug. 3. All the CCD images were corrected for bias, flat field, 
and cleaned of cosmic rays, using the IRAF package\footnote{IRAF, the Image Reduction and Analysis Facility, is 
distributed by the National Optical Astronomy Observatory, which is operated by the Association of Universities 
for Research in Astronomy(AURA), Inc. under cooperative agreement with the National Science Foundation(NSF).}. 
Since SN 2013en located within the host galaxy, special care is required  in the measurement of the 
instrumental magnitude of the supernova. Therefore, the template subtraction was applied by  using the LJT observations 
obtained on October 23.50 UT, 2014 when the SN faded away. The instrumental magnitude measured by subtraction were converted 
to the standard Johnson $BV$ \citep{John66} and Kron-Cousins $RI$ \citep{Cous81} systems through 
transformations established by observing \citep{Land92} standards during photometric nights. The averaged 
value of the photometric zeropoints, determined on two photometric nights, was used
to calibrate the local standard stars in the field of SN 2013en. Table~\ref{Tab:Ref} lists the standard 
$BVRI$ magnitudes and the corresponding uncertainties of nine local standard stars labeled in Fig.~\ref{Fig:1}. 
The magnitudes of these stars were then used to transform the instrumental magnitudes of SN 2013en to 
those of the standard $BVRI$ system. Table~\ref{Tab:LC} lists the final flux-calibrated magnitudes of SN 2013en.  A single-epoch $VRI$-photometry of SN~2013en 
obtained on 2013 Jul. 31 with the Asiago
1.82-m Copernico Telescope is also listed in Table~\ref{Tab:LC}.

\begin{figure}
\centering
\includegraphics[width=0.48\textwidth, angle=360]{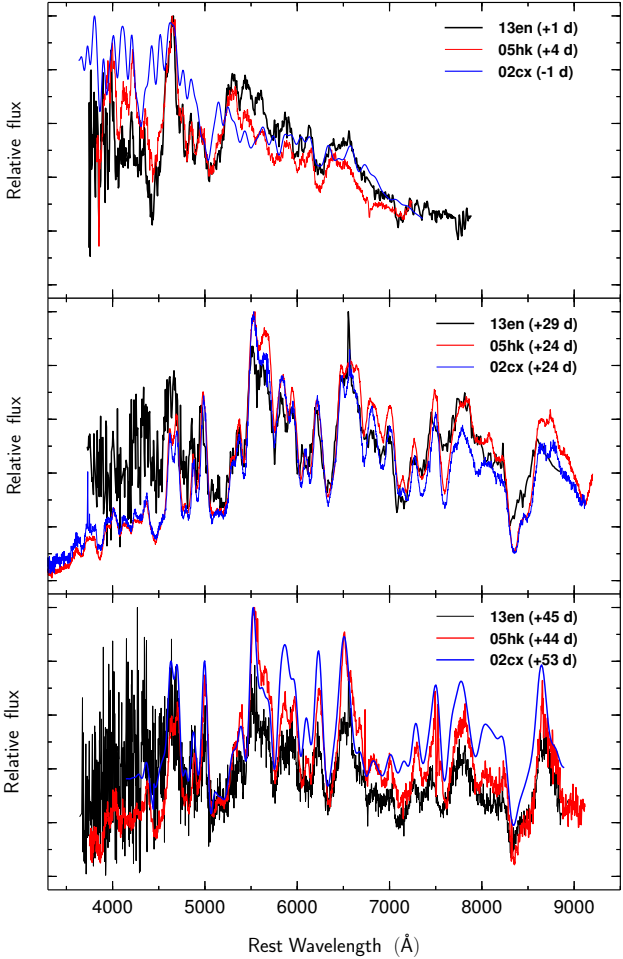}
\caption{Spectra of SN~2013en taken at +1 (black; top panel), +29 (black; middle
panel), and +45 days (black; bottom panel) relative to $R$-band maximum light. Overplotted are the  spectra of 
SN~2002cx (at $\sim$+10, +24, and +53 days in the top, middle, and bottom 
panels, respectively) and SN~2005hk (at $\sim$+4, +24, and +44 days in 
the top, middle, and bottom panels, respectively). 
All of the spectra have been corrected for the redshift
          of the host galaxy and for the reddening.
         }
\label{Fig:5}
\end{figure}

\begin{figure}
\centering
\includegraphics[width=0.45\textwidth, angle=360]{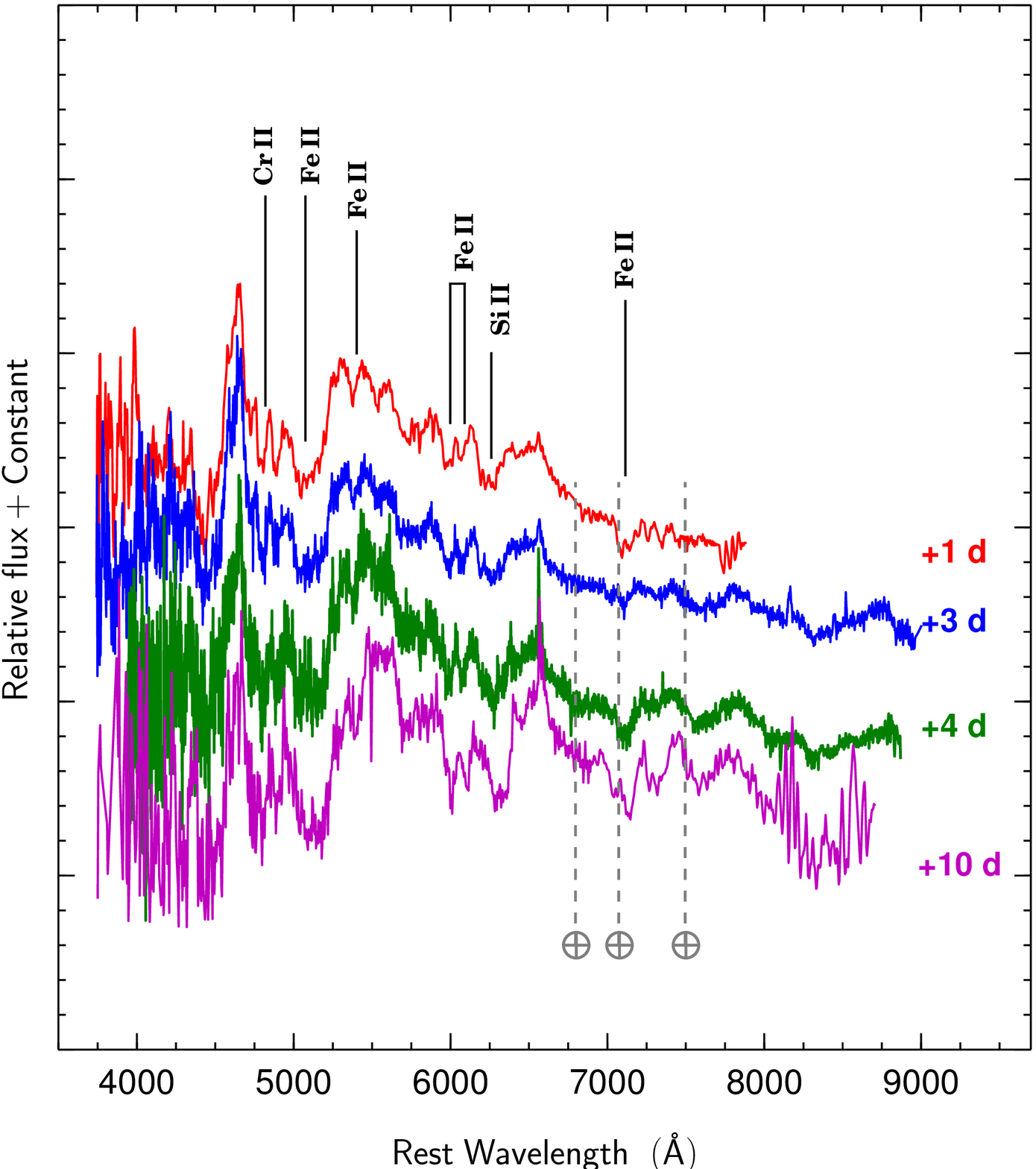}
    \vspace{0.2in}
\includegraphics[width=0.45\textwidth, angle=360]{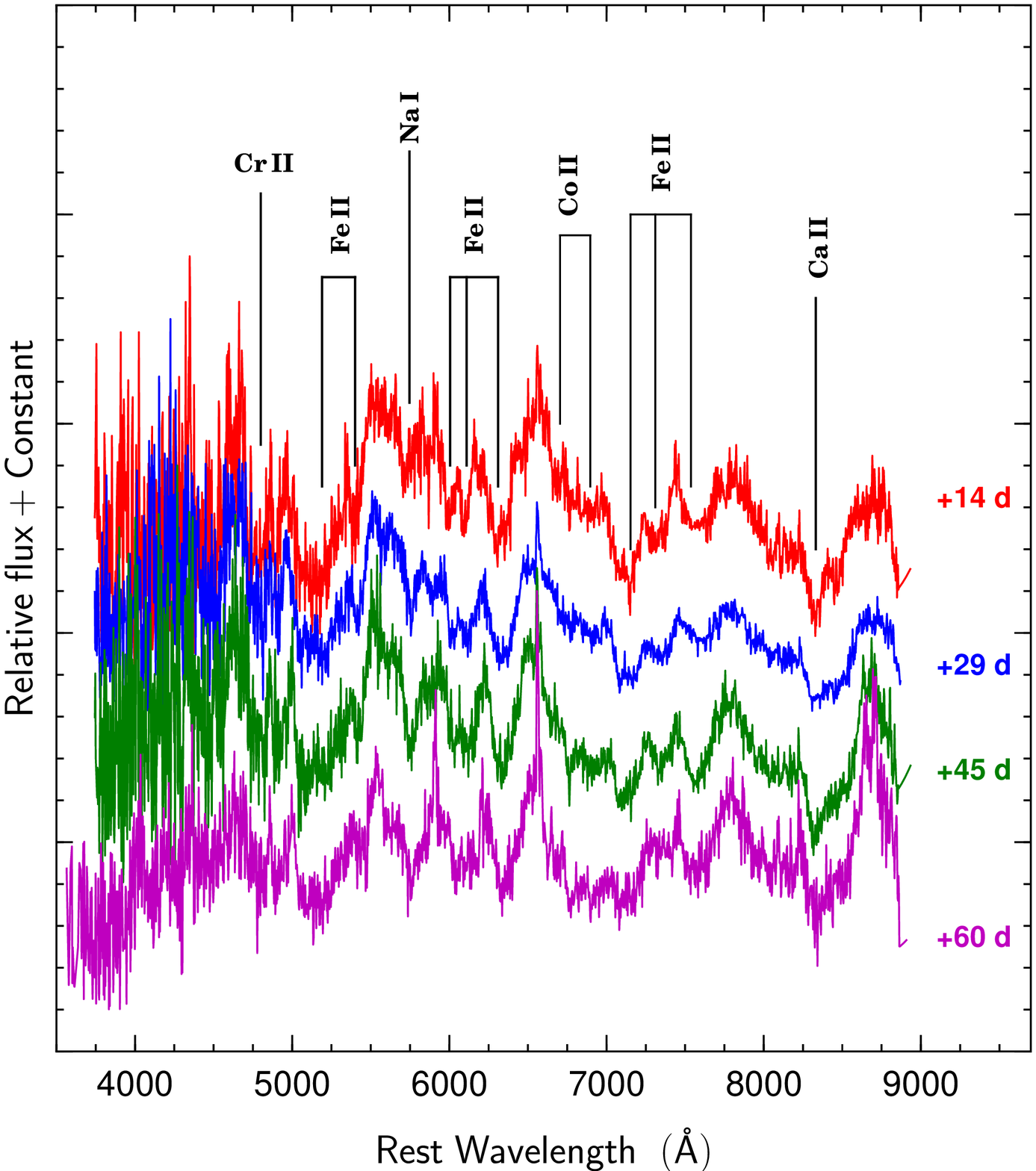}
\caption{ Spectral evolution of SN~2013en. The phases marked are relative to 
          date of $R$ maximum. The spectra have been corrected for the redshift
          of the host galaxy and for the reddening (assuming $E(B-V)_{\rm{tot}}\sim0.5\,\rm{mag}$). Line identification is 
          based on \citet{Bran04}. For clarity the spectra have been shifted vertically, residual 
          telluric absorption lines are marked with ``$\oplus$'' symbols.
         }
\label{Fig:6}
\end{figure}

\begin{table*}
\centering
\caption{The $BVRI$ Photometry of SN 2013en from LJT+YFOSC. }
 \begin{tabular}{@{}rcccccccl}
 \hline\hline
UT Date & MJD & phase$^a$  & $B$(mag) & $V$(mag) & $R$(mag) & $I$(mag) & Telescope & Observer \\
\hline
Jul. 31.87 & 56504.87   &       1.07    &       --              &       17.40(03)       &       16.74(04)       &       16.28(04)  & ACT+AFOSC  &L., Tomasella\\
Aug. 03.61 & 56507.61	&	3.81	&	18.03(05)	&	17.35(03)	&	16.82(03)	&	16.37(02)  & LJT+YFOSC	&J.J., Zhang\\
Aug. 15.73 & 56519.73	&	15.93	&	19.31(08)	&	17.89(04)	&	17.13(03)	&	16.55(03)  & LJT+YFOSC	&J.J., Zhang\\
Aug. 17.70 & 56521.70	&	17.90	&	19.42(08)	&	17.95(04)	&	17.22(03)	&	16.62(03)  & LJT+YFOSC	&J.J., Zhang\\
Sep. 13.63 & 56548.63	&	44.83	&	20.62(10)	&	18.91(04)	&	18.07(04)	&	17.61(03)  & LJT+YFOSC	&J.J., Zhang\\
Sep. 28.51 & 56563.51	&	59.71	&	20.77(10)	&	19.12(06)	&	18.42(05)	&	17.92(03)  & LJT+YFOSC	&J.J., Zhang\\
Oct. 09.50 & 56574.50	&	70.70	&	20.94(10)	&	19.28(06)	&	18.69(06)	&	18.19(05)  & LJT+YFOSC	&J.J., Zhang\\
Oct. 10.50 & 56575.50	&	71.70	&	21.05(10)	&	19.32(08)	&	18.71(07)	&	18.24(08)  & LJT+YFOSC	&J.J., Zhang\\
\hline
\end{tabular}

\medskip
\flushleft
Uncertainties, in units of 0.01 mag, are 1$\sigma$; MJD = JD-2400000.5.\\
$^a$ Relative to the date of $R$-band maximum (MJD =56503.80)
\label{Tab:LC}
\end{table*}

\begin{table*}
\centering
\caption{Journal of Spectroscopic Observations of SN 2013en. }
 \begin{tabular}{@{}rcccccccl@{}}
 \hline\hline
UT Date & MJD  &  Epoch$^{\ a}$ & Res & Range  &Airmass & Exp.time& Telescope&Observer \\
&(-240000.5) & (days) & (\AA/pix) & (\AA) &  & (sec) &(+Facility)& \\
\hline
 Jul. 31.85	&	56504.85	&	1.05	&	13.0	&	3800-8000	&	1.11	&	1800	&	ACT+AFOSC	&	L., Tomasella	\\
Aug. 02.61	&	56506.61	&	2.81	&	17.5	&	3500-8900	&	1.02	&	1800	&	LJT+YFOSC	&	 J.J., Zhang	\\
Aug. 03.73	&	56507.73	&	3.93	&	6.7	&	5000-9800	&	1.14	&	2700	&	LJT+YFOSC	&	 J.J., Zhang	\\
Aug. 03.76	&	56507.76	&	3.96	&	7.5	&	3800-7200	&	1.29	&	2700	&	LJT+YFOSC	&	 J.J., Zhang	\\
Aug. 09.60	&	56513.60	&	9.80	&	20.3	&	3500-8800	&	1.08	&	3600	&	XLT+BFOSC	&	X.F., Wang 	\\
Aug. 13.73	&	56517.73	&	13.93	&	17.5	&	3500-8900	&	1.29	&	1800	&	LJT+YFOSC	&	 J.J., Zhang	\\
Aug. 28.59	&	56532.59	&	28.79	&	17.5	&	3500-8900	&	1.01	&	2700	&	LJT+YFOSC	&	 J.J., Zhang	\\
Sep. 13.64	&	56549.14	&	45.34	&	17.5	&	3500-8900	&	1.01	&	3600	&	LJT+YFOSC	&	 J.J., Zhang	\\
Sep. 28.53	&	56563.53	&	59.73	&	17.5	&	3500-8900	&	1.04	&	3600	&	LJT+YFOSC	&	 J.J., Zhang	\\
\hline
\end{tabular}

\medskip
\flushleft
$^a$ Relative to the date of $R$-band maximum (MJD=56503.80)\\
\label{Tab:Sp}
\end{table*}

A journal of the spectroscopic observations of SN 2013en is given in Table \ref{Tab:Sp}.  A total of eight low 
resolution spectra obtained with the 2.4-m telescope (+YFOSC) at Lijiang Observatory (hereafter LJT), and the 2.16-m telescope (+BFOSC) at Xinglong 
Observatory (hereinafter XLT). A spectrum taken on 2013 Jul. 31 with the 1.82-m Copernico 
Telescope (+AFOSC) at Asiago Observatory (hereafter ACT) is also included in the analysis. All spectra were 
reduced using standard IRAF routines. Flux calibration of the spectra was performed by means of
spectrophotometric standard stars (i.e., LTT-1020) observed at similar air mass on the same night as the SN. 
With our own routines, the extracted, wavelength-calibrated spectra were corrected for continuum atmospheric extinction 
using mean extinction curves for Li Jiang and Xinglong Observatories. Additionally, telluric lines have been removed from the data.

\section{Analysis}
\label{sec:analysis}

\subsection{Light curves}
\label{sec:lc}

The host galaxy of SN~2013en, UGC~11369, is an SBa galaxy with a recession velocity of $4559\pm 35$\,$\rm{km\,s^{-1}}$ \citep{Falc99}.
Adopting a $\Lambda$-cold-dark-matter cosmological model with parameters $\rm{\Omega_{\Lambda}}=0.73$, $\rm{\Omega_{m}}=0.27$, $H_{0}=73\pm 5\ \rm{km\,s^{-1}\,Mpc^{-1}}$,  
the virgo infall distance of UGC 11369 is  $D=66.2\pm 4.7\,\rm{Mpc}$ ($\mu=34.11\pm 0.15\,\rm{mag}$) , which will 
be adopted in this work in the calculations of the luminosity of SN 2013en. The redshift of the UGC~11369 of 
$z\rm{=0.015207}$ is used as the redshift of SN~2013en.

The observed $BVRI$ light curves are presented in Figure~\ref{Fig:2}. 
Unfortunately, our photometry data of SN~2013en are quite sparse 
and the photometry observations began after the maximum light. 
Therefore, the maximum-light date of SN~2013en cannot be directly determined by 
only using our filtered data. To complement the early photometric 
points we combined the early-time unfiltered data of SN~2013en with our $R$--band light-curve
data which is shown in Figure~\ref{Fig:2}. The unfiltered images were obtained during automatic 
surveys at Monte Agliale Observatory in the framework of the ISSP\footnote{http://www.oama.it.}.
However, there might be possible differences between the ISSP unfiltered magnitudes and our R-band 
results. Fortunately, \citet{Zheng13} did a detailed comparison for unfiltered magnitudes from ISSP and the 
standard broad R-band magnitudes. They found that these magnitudes are comparable within an 
uncertainty of $5\%$  (see \citealt{Zheng13, Li03b}). Therefore, it is reasonable to combine the ISSP 
unfiltered and our standard R-band results.

For comparison the $BVRI$ light curves of SNe 2002cx (see \citealt{Li03}) and 
2005hk (which is a well-observed SN that is extremely 
similar to SN 2002cx, see \citealt{Phil07}) overplotted in Fig.~\ref{Fig:2}. 
The light curves of the comparison sample have been shifted in peak magnitudes 
and maximum-light time  to  match SN 2013en. 
As shown in Fig.~\ref{Fig:2}, the light curves of SN~2013en seem to follow
the photometric evolution seen in SNe~2002cx and 2005hk. In addition, it is shown that the SN~2013en 
likely peaked in the R-band at around 2013 July 30 (MJD~$56503.8\pm1.8$).

Fig.~\ref{Fig:3} shows that SN~2013en likely peaked at $\rm{m_{R}\sim16.7\pm0.2\,mag}$ in $R$-band. 
By adopting a distance modulus of $\mu=34.11\pm 0.15\,\rm{mag}$ and considering the total extinction of 
host galaxy and Milky Way (see Section~\ref{sec:spectra}), we obtain that SN~2013en 
peaked at $M_{\rm{R}}\approx -18.6\,\rm{mag}$ (with a total uncertainty of $\sim$0.3--0.4\,$\rm{mag}$), comparable 
to SN~2002cx ($M_{\rm{R}}\approx -17.6\,\rm{mag}$, \citealt{Li03}) and 
SN~2005hk ($M_{\rm{R}}\approx -18.3\,\rm{mag}$, see \citealt{Phil07}). The uncertainty in 
this measurement is relatively large due to poorly-sampled light curves at around the maximum light
and uncertainties in distance and extinction correction.

\subsection{Spectra}
\label{sec:spectra}

\begin{figure}
\centering
\includegraphics[width=0.45\textwidth, angle=360]{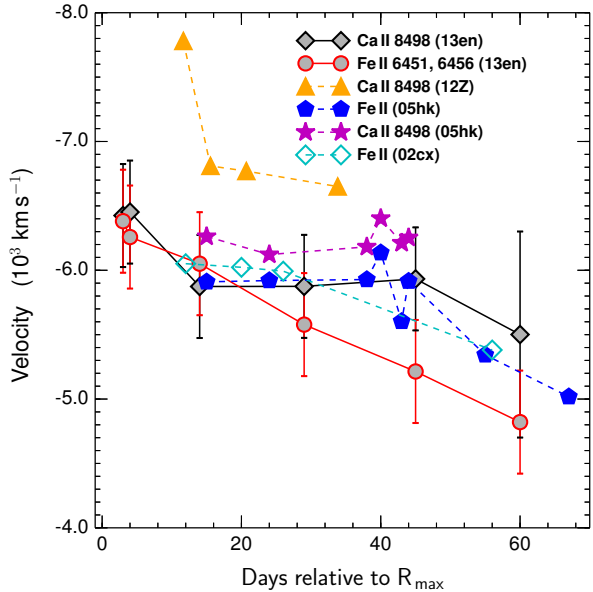}
\caption{Expansion-velocity evolutions of \Caii\ and \Feii\ lines as measured from the spectra of SN~2013en.
         For a comparison, temporal evolutions of similar lines of SN~2002cx, SN~2005hk and SN~2012Z are also plotted.}
\label{Fig:7}
\end{figure}

Figure~\ref{Fig:3} shows time series of our spectra covering the evolution of SN 2013en from 
around maximum light to two months after maximum light. Note that the spectrum at $t\approx+4$ days was compounded by 
the spectra obtained by the blue and red grisms. 
It is shown that the spectra of SN~2013en remain almost unchanged
during a period from day +14 to day +29 and the spectrum of day +60 is very similar to that of day +45.
The 2013en spectrum exhibits strong \Nai\ D interstellar
absorption from both the Galactic and host-galaxy components (see the inset of Fig~\ref{Fig:3}). The equivalent width (EW) of \Nai\ D absorption 
is suggested to be correlated with the  line-of-sight reddening, but with a large scatter (e.g., \citealt{Tura03, Pozn12}).
The \Nai\ D absorption due to the host galaxy of SN 2013en is measured to be about {1.62\AA} at $t\approx+4$ days, which correspond to a possible
host galaxy reddening $E(B-V)_{\rm{host}}=0.26\pm0.03\,\rm{mag}$ \citep{Tura03, Pozn12}. 
The Milky Way reddening is $E(B-V)_{\rm{Gal}}=0.254\,\rm{mag}$ according to \citet{Schl98}. 
Therefore, a total reddening (host galaxy+Milky way) of $E(B-V)_{\rm{tot}}\approx0.5\,\rm{mag}$ is adopted 
for SN~2013en in this paper.  And a reddening law with $\rm{R_{V} = 3.1}$ of \citet{Card89} is assumed 
for the extinction correction.

In Figure~\ref{Fig:4}, we compare the spectra SN~2013en wit those of SN~2002cx and SN~2005hk (02cx-like), SN~1999ac (a 1991T-like) and 
SN~2003du (a normal SN Ia) at similar phases. 
No signatures of hydrogen and helium are found in spectra of SN~2013en, which indicates that its progenitor  was likely
an evolved star \citep{Fole12, Fole13} or an envelope-stripped star like the Wolf–Rayet star. One can see that the spectra of SN~2013en and their evolution are quite 
different from those of normal SNe Ia, but are very similar to SN~2002cx and SN~2005hk \citep{Li03, Bran04}.

A detailed spectra comparison between SN~2013en and SN 2002cx, SN 2005hk \citep{Li03, Bran04, Jha06, Phil07} is performed  at $t=+1$, +29, and +45 days after maximum 
brightness ($R$--band), as shown 
in Fig.~\ref{Fig:5}. It is found that they are nearly identical, and the line features in SN~2013en spectra 
closely resemble those of SN~2002cx and SN~2005hk, except for a slight 
difference in line velocities and widths. Furthermore, the evolution of some typical line 
features seen in the spectra of SN~2013en are shown in Figure~\ref{Fig:6}. As it can be seen that the spectra of SN~2013en are dominated by \Feii\ lines. Also, \Coii,
\Nai, and \Caii\ lines are clearly visible in the spectra.  A detailed 
line identification for the 2002cx-like SNe was addressed in \citet{Bran04}.

\begin{figure}
\centering
\includegraphics[width=0.45\textwidth, angle=360]{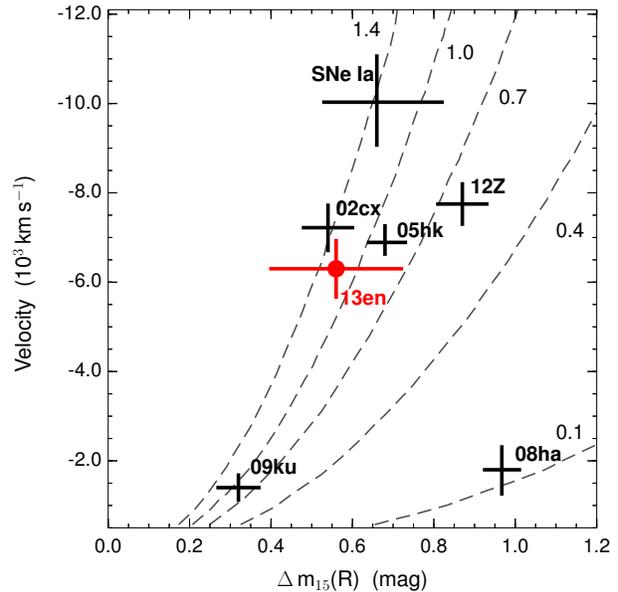}
\caption{Comparison of $\Delta\,m_{15}(R)$ and the photospheric velocity at around the maximum light ($t\approx+3$~days) 
         for various  objects in the SN 2002cx-like class. The dashed lines represents the relationship between ejecta 
        velocity and $\Delta\,m_{15}$ for (from bottom right to top left) 0.1, 0.4, 0.7, 1.0, 
        and $1.4\,\rm{M_{\odot}}$ of ejecta mass. SN 2013en is marked with a red filled circle.
         }
\label{Fig:8}
\end{figure}

\subsection{Ejecta Velocity}
\label{sec:velocity}

The blueshift of the absorption lines in SN spectra has been used as an indicator of expansion velocity,  and 
they can provide information about the kinetic energy of the SN explosion, the chemical 
stratification of the ejecta \citep{Parr14}, and even the progenitor environment \citep{Wang13}. Here, the ejecta velocity of SN 2013en is derived from the absorption 
features of \Caii\ and \Feii\ lines. The location of the blue-shifted absorption minimum is 
measured by using the direct measurement of absorption minima. Fig.~\ref{Fig:7} shows that 
the ejecta velocity of SN~2013en measured from the minimum blueshift of \Caii\ and \Feii\ evolves 
from $\simeq6400\,\rm{km\,s^{-1}}$ at $t=+3$ days to $\simeq5500\,\rm{km\,s^{-1}}$ at $t=+60$ days. For a 
comparison, the ejecta velocities of SN~2002cx, SN~2005hk \citep{Phil07} and SN~2012Z \citep{Stri15} are also plotted in Fig.~\ref{Fig:7}. 

Based on the unfiltered and R-band light curve presented in Figure~\ref{Fig:2}, we measured for SN 2013en the decline 
rate within the first 15 days after the maximum, $\Delta m_{15}(R)=0.55\pm 0.18\ \rm{mag}$. 
The large error is due to a relatively large uncertainty in estimating the date of maximum light (see Section~\ref{sec:lc}). 
Using Arnett's Law \citep{Arne82}, we can derive a ejecta mass-dependent relationship between ejecta velocity
and $\Delta m_{15}(R)$ (see also \citealt{Nara11}). In Fig.~\ref{Fig:8}, we present such a relation for SN~2013en 
and other comparison sample of 02cx-like SNe Ia. It can be seen that the expansion velocity of SN~2013en is significantly 
slower than that of normal SNe Ia, but the location of SN~2013en in the plot is consistent with that of SN~2002cx 
and SN~2005hk. \citet{Nara11} suggested that the peak luminosities of 02cx-like SNe Ia
is related to their ejecta velocities, except for SN~2009ku. It is obvious that SN 2013en is 
in line with such a relation.

\section{Discussion and Conclusion}
\label{sec:discussion}

In this paper, we present optical photometry and spectroscopy of a 02cx-like peculiar supernova
SN~2013en. Combining our light curves and  the early-time unfiltered data, 
we were able to determine that SN~2013en likely peaked in R-band at around 2013 July 30 ($\rm{MJD=56503.80}$), 
with $m_{R}\sim16.7\pm0.2\,\rm{mag}$ and $\Delta m_{15}(R)=0.55\pm 0.18\ \rm{mag}$.

Based on the \Nai\ D absorption from the spectra, we estimated that SN 2013en 
suffered a total reddening of $E(B-V)_{\rm{tot}}\approx0.5\,\rm{mag}$. Assuming a  
distance modulus $\mu=34.11\pm 0.15\,\rm{mag}$, we found that SN 2013en has a 
absolute peak magnitude of $\rm{-18.6\,mag}$ in R-band.  The brightness of SN~2013en is 
consistent with that of peculiar subluminous SN~2002cx and SN~2005hk.

Besides the photometric evolution, the spectra of SN~2013en are found to show similar features 
and evolution to those SNe~2002cx and 2005hk. These results indicate that SN~2013en is probably produced from the same progenitor 
scenario as SNe~2002cx and 2005hk. However, the exact nature of progenitor systems of SNe Iax remains 
unclear. The HST analysis for one SN Iax SN~2012Z suggests that a binary system consists of 
a He-star companion and a CO WD is likely to be its progenitor system \citep{McCu14}. 
However, the observations for other SN Iax objects also show that a wide variety of 
progenitor scenario may be realized within the SNe Iax class \citep{Fole13, Fole14, Stri15, Fole15}.

Theoretically, as mentioned previously, numerical simulations show that weak deflagrations of 
the Ch-mass CO WDs produce observational characteristics of 2002cx-like SNe Iax quite well \citep{Jord12, Krom13, Fink13}, 
although it appears difficulties in explaining 2008ha-like SNe Iax (SNe~2008ha 
and 2010ae, see \citealt{Vale08, Stri14}) which have a very low $^{56}\rm{Ni}$ mass of $\sim0.003\,\rm{M_{\odot}}$. 
Recently, however, hybrid C-O-Ne Ch-mass WDs have also been suggested
likely to trigger weak pure deflagration explosions of SNe Iax \citep{Deni15}. Because the Hybrid 
WDs have much lower C to O abundance 
ratios at the moment of the explosive C ignition than their pure CO counterparts \citep{Deni15}, 
which probably will lead to a low $^{56}\rm{Ni}$ mass after the SN explosion. 
Recent hydrodynamical simulations for the weak pure deflagration explosions of a Ch-mass hybrid WD have 
shown that this specific explosion can explain the 
observational features of faint SNe Iax such as SN~2008ha (see \citealt{Krom15}). 
On the other hand, the fallback core-collapse explosions of massive stars were also proposed for the peculiar SN Iax 
SN~2008ha (see \citealt{Vale08, Mori10, Fole10}). Nevertheless, to date, weak deflagration explosions of a Ch-mass WD 
seem to give better explanations for the observational properties of SNe Iax.

\section*{Acknowledgments}

We acknowledge the anonymous referee for valuable comments that helped us to improve the paper. 
We acknowledge the support of the staff of Li-Jiang 2.4-m telescope (LJT) and Xinlong 2.16-m telescope. 
Funding for the LJT has been provided by Chinese academe of science (CAS) and the People's Government of Yunnan 
Province. Partially based on data obtained in Asiago, by the Copernico 1.82-m telescope operated by INAF OAPd. 
Z.W.L is supported by the Alexander von Humboldt Foundation. J.J.Z is supported by the National Natural 
Science Foundation of China (NSFC, grant 11403096). L.T is partially supported by the PRIN-INAF 2011 with 
the project ``Transient Universe: from ESO Large to PESSTO''. X.F.W is supported by the Major State Basic 
Research Development Program (2013CB834903), the NSFC (grants 11073013, 11178003, 11325313), Tsinghua 
University Initiative Scientific Research Program, and the Strategic Priority Research Program ``The 
Emergence of Cosmological Structures'' of the Chinese Academy of Sciences (grant No. XDB09000000).

\bibliographystyle{mn2e}

\bibliography{ref}

\end{document}